\documentclass[nohyper,proceedings]{JHEP}
\usepackage{epsfig}                   
\usepackage{psfig}

\conference{Nonperturbative Quantum Effects 2000 }

\conference{}
\makeatletter
\gdef\@maketitle{
\begin{flushleft}
\vskip-5em
{\em
\parbox[t]{.6\linewidth}{\small 
\leftline{Talk presented by GMTW at TMR-conference "Nonperturbative
Quantum Effects 2000"} 
\leftline{Typeset in JHEP proceedings style}
}
\hfill
\parbox[t]{.35\linewidth}{\small
\rightline{KCL-MTH-00-54}
\rightline{hep-th/0010082}
}
}
\vskip -1.5em
\null
    \vskip 1.5em plus .4fil			
        {\LARGE \sffamily                       
        \bfseries                               %
        \@title\par}                            %
    \vskip .6em plus .06fil minus .5ex          %
    \rule\textwidth{\if@proc\else1.5\fi\p@}     
    \vskip 1em plus .06fil minus .6ex           
        {\normalsize \bfseries \sffamily        %
        \@author \par}                          
    \vskip 2em plus 0.05fil minus 1ex           %
        \parbox\textwidth{\unhbox\abstract@box} 
    \vskip 2em plus 1em minus 1ex		%
        \par                                    %
    \@keywords                                  
        \par
    \vskip\baselineskip
        \dedic@box                              
  \end{flushleft}
}
\makeatother
\newcommand{\be}{\begin{equation}}
\newcommand{\ee}{\end{equation}}

\newcommand{\cev}[1]{\langle \,#1\,|}
\newcommand{\cg}{{\cal G}}

\newcommand{\clim}{{\ensuremath{c\to 1}}}

\newcommand{\D}{{{\rm d}}}
\newcommand{\ds}{\displaystyle}

\newcommand{\gir}{g_{\hbox{\tiny IR}}^{\vphantom\phi}}
\newcommand{\guv}{g_{\hbox{\tiny UV}}^{\vphantom\phi}}

\newcommand{\One}{{\hbox{{\rm 1{\hbox to 1.5pt{\hss\rm1}}}}}}
\renewcommand{\One}{{\mathbb 1}}
\renewcommand{\One}{{\rm 1\!\!1}}

\renewcommand{\vec}[1]{|\,#1\,\rangle}

\title{Renormalisation group flows of boundary theories}

\author{
K. Graham, 
I. Runkel\thanks{%
Address from October 2000:
LPTHE, Paris VI}~
and G.M.T. Watts  \\

Department of Mathematics,
King's College London, 
Strand, London WC2R 2LS, U.K.
\\
E-mail:
{\tt
kgraham@mth.kcl.ac.uk, 
ingo@lpthe.jussieu.fr,
gmtw@mth.kcl.ac.uk
}\\
}
\abstract{          
We review recent developments in the theory of renormalisation group
flows in minimal models with boundaries.
Among these, we discuss in particular the perturbative calculations of
Recknagel et al, not only as a tool to predict the IR endpoints of
certain flows, but also as a motivation for considering the particular
limiting case of $c\,{=}\,1$. 
By treating this limit, we are able to investigate a wide class of
perturbations by considering them as deformations away from the 
$c\,{=}\,1$ point. 
We also present the truncated conformal space approach as a tool for
investigating the space of RG flows and checking particular predictions.
}
\begin{document}
\section{Introduction}

In this talk we consider two dimensional theories defined on manifolds
with boundaries. This allows the possibility for renormalisation group
(RG) flows generated by boundary fields and flows between different
boundary conditions.
The discussion of RG flows of boundary
theories can be usefully divided into 

\begin{enumerate}
\item {Purely boundary flows}

\item {Combined bulk and boundary flows}
\end{enumerate}

In this talk we shall consider only the first class of flows.
Some examples of combined bulk and boundary flows are considered
in the talk by R.~Tateo \cite{tateo-talk}.
Before starting our discussion of boundary flows, it is useful to
recall the well-known results for purely bulk perturbations.

\section{The bulk case}


In two-dimensions, the field
theory at an RG fixed point is a conformal field theory (cft).
Hence, near such a fixed point, the (bulk) action takes the form
\be
  S_{\mathrm{bulk} }
= S_0  
+ \sum_i \lambda_i \int\! \varphi_i(z,\bar z)\; \D^2 z
\;,
\label{eq:sbulk}
\ee
where $S_0$ is the fixed point cft action and $\varphi_i(z,\bar z)$
are conformal scaling fields of scaling dimension $x_i$, with
corresponding couplings $\lambda_i$.  
Under scale transformations, the couplings have $\beta$--functions
\be
  \dot{\lambda}_i 
\equiv \beta_i
 = (2 - x_i)\lambda_i
 + \ldots
\;,
\label{eq:beta1}
\ee
so that $\lambda_i$ increases or decreases as $x_i\,{<}\,2$ or $x_i\,{>}\,2$
respectively, as in figure \ref{fig:nbhd}.
(The case $x_i\,{=}\,2$ needs more careful analysis to decide the nature of
the corresponding RG flow, if any)

\FIGURE{
\parbox{5cm}{%
\input{nbhd.tex}
}
\caption{\newline The RG flows in the neighbourhood of a fixed point}
\label{fig:nbhd}
}


Order is brought to the space of RG flows by 
Zamolodchikov's $c$--theorem \cite{Zamo-LG}.
This states that for unitary theories there is a function $C$ which is  
decreasing along RG flows and which is equal at a fixed point to the
conformal central charge of the corresponding cft.
In terms of the trace of the stress-energy tensor one has
\[
\cases{
  \dot C 
= -  
     \frac{3}{4}
  \, \langle 
  \, \Theta(1) 
  \, \Theta(0) 
  \, \rangle
\cr
  C \Big|_{\Theta=0} = c \;,\;\; \hbox{central charge}
}
\]
As a result, one knows that the bulk RG flows always cause $c$ to
decrease and one has a schematic picture as in figure
\ref{fig:flows}. 
Flows which have non-trivial cfts as both UV and IR fixed points are
known as `massless flows', while those which end at a fixed point with
$c\,{=}\,0$ are known as `massive flows'. In the latter case these define
massive scattering theories in the neighbourhood of the IR fixed
point. 
\FIGURE{%
\noindent\parbox{7cm}{%
\input{flows.tex}
\caption{\newline{}Schematic picture of the space of bulk RG flows}
\label{fig:flows}
}}

Given this picture, the next most important fact is that for many
cases one has a classification of the conformal field theories, for
example for unitary minimal models with $c\,{<}\,1$ one has the `ADE'
classification of Cappelli et al \cite{CIZ87}
(for these and other results on the minimal models see
e.g. \cite{Ybk}).
From this result one knows the possible UV and IR endpoints, and all
the possible relevant $(x_i\,{<}\,2)$ operators and one
has then `simply' to join up the cfts by flows.

Finally, one has the Landau-Ginzburg model for the theories with a
given UV fixed point \cite{Zamo-LG}
in which the space of relevant perturbations are described by an
action for scalar field(s) with a particular class of polynomial
potentials. This picture gives one a heuristic understanding of the
space of RG flows, and the vacuum structure of the IR fixed points;
for a detailed treatment of the tricritical Ising model using this
method, see \cite{LMCa1}.

Let's now turn to the case of models with boundaries and see how much
can be said there.



\section{The boundary case}

If the space on which our theory is defined has boundaries, then we
must consider the possible boundary conditions that can be put on
these boundaries and the space of fields which can exist on
corresponding boundary conditions.
For a semi-infinite cylinder, a boundary condition $\alpha$ on the end
of the cylinder defines a `boundary state' $\cev{B_\alpha}$. 
One can take the inner product of such a state with a bulk states
$\vec\psi$ (as in figure \ref{fig:cyl}), but the boundary state is not
actually a normalisable state in the bulk Hilbert space.

\FIGURE{
\parbox{5cm}{
\input{cyl.tex}
\caption{\newline{}Bulk and boundary states associated to a cylinder}
\label{fig:cyl}
}}

If the bulk theory is at an RG fixed point, it is not necessarily the
case that the boundary is invariant under scale transformations.
Those boundary conditions which are invariant are called `conformal
boundary conditions' and Cardy gave the conditions for a boundary
state to correspond to a conformal boundary condition in 
\cite{Card4}.
For each bulk CFT there is a set of possible conformal boundary
conditions
$\{\, B_\alpha\, \}$, and associated to each conformal boundary
condition there is a set of conformal scaling fields which can exist
on that boundary.

Consequently, 
near a fixed point the action for the theory on a space $M$ with
boundary $\partial M$ takes the form
\be
  S 
= S_{\rm bulk}
+ \sum_i \mu_i     \oint_{\partial M}\! \phi_i(l) \, \D l
\;,
\label{eq:sbdy}
\ee
where $S_{\rm bulk}$ is the action (\ref{eq:sbulk}) and $\phi_i$ are
boundary scaling fields of dimension $h_i$.


For each CFT and boundary condition $B_\alpha$ we have to determine
the possible RG trajectories  generated by both bulk and boundary
perturbations, and the IR bulk theory and its boundary condition
$B'$ which is quite a challenge.

There is one important simplification that we can make, and that is to
set all the couplings $\lambda_i$ in the bulk action (\ref{eq:sbulk})
to zero. 
If that is the case, then (to all orders in perturbation theory) the
bulk theory remains conformally invariant, and so the cft of the IR
fixed point will be the same as that at the UV fixed point.
Consequently, the only effect of the RG is to flow in the space of
possible boundary conditions, in which the fixed points are the
conformal boundary conditions.

We can calculate the $\beta$ function
for the coupling $\mu_i$ to find
\be
  \dot \mu_i
= (1-h_i) \mu_i \;+\; \ldots
\;.
\label{eq:beta2}
\ee
Again we have a flow {\em out} of a UV fixed point for $h_i \leq 1$
and a flow {\em into} an IR fixed point for $h_i \geq 1$.
Hence to understand the space of flows starting at a given boundary
condition, we only need to consider the perturbations by operators
with weight $h_i\,{<}\,1$, which (for a rational cft, and in
particular for a minimal model) will be a finite
dimensional space of perturbations.


How do we bring order to this space of flows?
The first requirement is to a have a quantitative estimate of the
number of degrees of freedom associated to a boundary condition (which
should then decrease along RG flows).

In \cite{AL91}, Affleck and Ludwig defined the `generalised ground
state degeneracy' $g_\alpha$ associated to a conformal boundary
condition $B_\alpha$ as follows.
Consider the partition function of a cft on a cylinder of length $R$
and circumference $L$ and with boundary conditions $\alpha$ and $\beta$
at the two ends, as in figure \ref{fig:cyl2}.

\FIGURE{
\parbox{5cm}{
\input{cyl2.tex}
\caption{The cylinder partition function}
\label{fig:cyl2}
}}

In terms of the Hamiltonian $H_{\alpha\beta}(R)$ propagating around
the cylinder, this partition functions has a canonical normalisation
\be
  Z_{\alpha\beta} 
\;=\;
  {\rm Tr}\,
  e^{-L\, H_{\alpha\beta}(R)}
\;.
\label{eq:Z1}
\ee
In the limit $R\to\infty$, one finds   
\be
  Z_{\alpha\beta} 
\;\sim\;
  g_\alpha\, g_\beta\, e^{-R\, E_0(L)}
\;+\;\ldots
\;,
\label{eq:g0}
\ee
where $E_0(L)$ is the energy of the ground state $\vec\Omega$ of  the
Hamiltonian propagating along the cylinder, and where the constants  
$g_\alpha$, $g_\beta$ are given by
\be
  g_\alpha
= \langle B_\alpha | \Omega \rangle
\;.
\label{eq:g1}
\ee
These constants are the generalised ground state degeneracies
associated to the conformal boundary conditions%
\footnote{
As pointed out by A.~Cappelli and W.~Nahm in the talk, care must be
taken when dealing with irrational theories for which it may be the
case that only {\em ratios} of the constants $g_\alpha$ can be defined
sensibly}.


For a perturbed theory with a non-trivial scale-dependence, it is not
so easy to define such ground state degeneracies. If we try to adapt
the previous method, instead of (\ref{eq:g0}), we find
\be
  Z_{\alpha\beta} 
\sim
  \cg_\alpha(L)\, \cg_\beta(L)\, e^{-R\, E_0(L)}
\;+\;\ldots
\;,
\label{eq:g2}
\ee
where the functions $\cg_\alpha(L)$ generically behave for large $L$
as  
\be
  \log \cg_\alpha(L)
= - L\, f_\alpha^B 
\;+\;
  g_\alpha(L)
\;.
\label{eq:g3}
\ee
Here $f_\alpha^B$ is the boundary free energy per unit length, and 
it is the function $g_\alpha(L)$ which interpolates ${\guv}$ and
${\gir}$.
If we try to calculate $\cg_\alpha(L)$ or $g_\alpha(L)$, we find
different behaviour for $h\,{<}\,1/2$ and $h\,{>}\, 1/2$.
(The special case $h{=}1/2$ is again different).

For $h\,{<}\,1/2$, $f_\alpha^B$ is physical and non-per\-turba\-tive.
This means that it is not possible to calculate $g_\alpha(L)$ in
perturbation theory.

For $h\,{>}\,1/2$, $f_\alpha^B$ is non-physical and divergent.
Surprisingly, this is a better situation than the previous one, as 
it is possible to remove the divergent terms proportional to $L$ and
calculate $g_\alpha(\mu_R)$  as a perturbation expansion in the
renormalised coupling $\mu_R$, as shown by Affleck and Ludwig in
\cite{AL93}. 

In this second case, Affleck and Ludwig show that
$g_\alpha$ defined via (\ref{eq:g3}) 
satisfies
\be
  \dot g_\alpha < 0
\ee
{\em to leading order in perturbation theory};
nevertheless this result has gone by the name of the ``$g$--theorem'',
namely that for a perturbation of a unitary theory by purely boundary
fields, the UV and IR boundary conditions satisfy ${\gir} \,{<}\, {\guv}$.
The upshot of this conjecture is that we can order our flows by the
value of $g$. 

Hence, in almost all respects the quantity $g_\alpha$ behaves for
boundaries as the central charge $c$ does for bulk theories -- except
that there is no non-perturba\-tive proof of 
the corresponding ``$g$--theorem''.
(For a recent development see \cite{new})

While this result was known for a long time, as were the values of
$g_\alpha$ for a large class of boundary conditions, it was not until
very recently that there was a complete classification of the possible
boundary conditions of the unitary minimal models.
This was found by Behrend et al. and discussed in a series of papers
\cite{BPZu1,BPPZs}.
They classified the boundary conditions which admit only a single
scalar $(h\,{=}\,0)$ boundary field -- equivalently, those for which
boundary correlation functions which obey the cluster property -- the
so--called `Cardy' boundary conditions. 

For the `A' type `diagonal' modular invariant theories, they found
that the elementary (Cardy type) boundary conditions are in 1--1
correspondence with the Virasoro highest weight representations
\be
   B_\alpha 
\; \leftrightarrow 
\; h_\alpha
\;.
\label{eq:Cbc}
\ee
As an example, consider the tricritical Ising model,
which is the `A' type unitary minimal model $M_{4,5}$. 
This cft has $c\,{=}\,7/10$ and six Virasoro representations of
interest; consequently there are six conformal boundary conditions.
It can be thought of as the continuum limit of a critical lattice
model on each site of which  there can either be a spin in state down
($-$) or state up ($+$), or the site can be empty ($0$).
The six Cardy boundary conditions can be labelled by the possible
values of the sites on the boundary and associated to the six
representations of the Virasoro algebra as in table \ref{tab:tcim}.

\TABLE{
\parbox{5.7cm}{\renewcommand{\arraystretch}{1.3}
$
\begin{array}{|cc|c|}
\hline
  h_{11} = 0   ~&~ h_{14} = 3/2  ~&~ h_{21} = 3/16 
\\

  - & + & 0 
\\[-5mm]

  \multicolumn{2}{|c|}{\underbrace{~~~~~~~~~~~~~~~~~~~~}}
& \\

  \multicolumn{2}{|c|}{g=0.5127}
& g=0.725

\\
\hline

  h_{12} = 1/10 ~&~  h_{13} = 3/5 ~&~ h_{22} = 3/80 
\\

  -0 & 0+ &  \hbox{`$d$'}
\\[-5mm]

  \multicolumn{2}{|c|}{\underbrace{~~~~~~~~~~~~~~~~~~~~}} 
& \\

  \multicolumn{2}{|c|}{g=0.8296}
& g=1.173

\\
\hline
\end{array}
$
}
\caption{%
The conformal boundary conditions of the tricritical Ising model}
\label{tab:tcim}
}

In \cite{Affl1}, the boundary condition `0' was identified as the free
boundary condition and `$d$' as an additional `degenerate' boundary
condition; however in view of the fact that `$d$' has the maximal
number of boundary degrees of freedom, it seems natural to consider it
as the free (or possibly freest) boundary condition.

The space of boundary RG flows in the tricritical Ising model has
been given recently by Affleck in \cite{Affl1}, which we reproduce in
figure \ref{fig:m45}.


\FIGURE{%
\noindent\parbox{7cm}{%
\input{m45.tex}
\caption{The boundary RG flows of the tricritical Ising model}
\label{fig:m45}
}}

A few comments are in order on the label `integrable' applied to the
flows generated by the fields $\phi_{(1,2)}$ and $\phi_{(1,3)}$. 
As discussed in \cite{GZam1}, one can show show the conservation 
{\em to first order in perturbation theory}
of non-trivial conserved quantities for perturbations by boundary
fields for which a similar argument shows the corresponding bulk
perturbation to be integrable. To be explicit, for a generic minimal
model this comprises perturbations by fields of type $(1,2)$, $(1,3)$
and $(1,5)$, and their duals $(2,1)$, $(3,1)$ and $(5,1)$. 
Further evidence for the integrability of the $(1,3)$ perturbation on
boundary conditions of type $B_{(1,s)}$ is given by the existence of
exact TBA equations which are supposed to describe these flows
\cite{LSS}. 
For the perturbation by $\phi_{(1,2)}\equiv\phi_{(3,3)}$
the obvious line crossings in figure \ref{fig:Graph1b12.45}
are usually taken as a clear indication of integrability of the RG flow.


What tools are there which would enable us to map out such a picture?
Here we list three:

\begin{enumerate}

\item
If $y\equiv1-h$ is small, then we can use perturbation theory as
outlined in \cite{AL91}.
In this way one can calculate $\Delta g$ along a flow.

\item

If the perturbation is integrable, one may be able to use exact
integral equation (TBA or NLIE) techniques.
In this way one may be able to calculate $g$, or some or all of the
spectrum, along the flow. 

\item

For any perturbation one can use the Truncated conformal space
approach (TCSA).
In principle, this can be used to calculate any quantity which can be
defined in terms of the perturbed conformal field theory construction.
However, one has little control over the accuracy of the results and
ranges of parameters for which the results are reliable.

\end{enumerate}

\noindent We now consider these in more detail.


\section{Perturbation theory}

Let us consider a perturbation by a single 
relevant ($h\,{<}\,1$) boundary field $\phi$.
The first condition we must check is that no new perturbations are
generated as counterterms to remove divergences in the
perturbation expansion in $\mu$.
No new counterterms are needed if the operator product
expansion of this field with itself takes the form
\begin{eqnarray}
  \phi(x)\; \phi(x')
= \frac{1}{(x{-}x')^{2h}}
\;&+&\;
  \frac{C\,\phi(x')}{(x{-}x')^h}
\;+\;
  \ldots
\;,
\nonumber
\\
&& ~~~~
  x>x'
\;,
\label{eq:ope}
\end{eqnarray}
where the right hand side contains no further contributions from
fields of weight $h'$ with $h'{\leq} 1{-}2y$.
Under this condition, Affleck and Ludwig found the $\beta$--function
to be (in a particular regularisation scheme)
\be
  \dot\mu
= y\,\mu
\;-\;
  C\,\mu^2
\;+\;
  \ldots
\;,
\label{eq:beta3}
\ee
where it is assumed that $\mu$ is $O(y)$ and terms of order $y^3$ have
been dropped; 
they also found the change in $g$ along the flow to be
\be
  \log\left(
  \frac{ g(\mu) }{ {\guv} }
  \right)
= - {\pi^2}y\mu^2
 \;+\;
    \frac{2\pi^2}{3}C\mu^3
\;+\;
  O(y^4)
\;.
\label{eq:dg0}
\ee
In figure \ref{fig:beta} we plot the $\beta$--function and the RG
flows, and find a perturbative fixed point at 
\newline
$\mu_R^* \,{=}\,y/C$. This gives
\be
  \log\left(
  \frac{ {\gir} }{ {\guv} }
  \right)
= - \frac{\pi^2}3\,
    \frac{y^3}{C^2}
\;+\;
  O(y^4)
\;.
\label{eq:dg}
\ee

\FIGURE{
\parbox{5cm}{
\begin{picture}(0,0)%
\epsfig{file=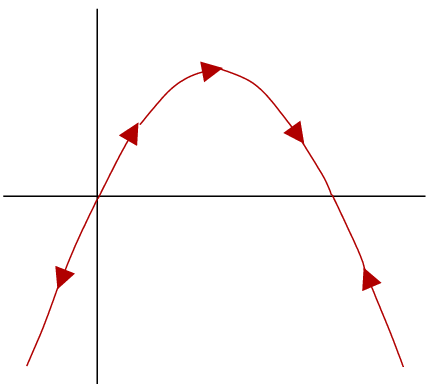}%
\end{picture}%
\setlength{\unitlength}{2960sp}%
\begingroup\makeatletter\ifx\SetFigFont\undefined%
\gdef\SetFigFont#1#2#3#4#5{%
  \reset@font\fontsize{#1}{#2pt}%
  \fontfamily{#3}\fontseries{#4}\fontshape{#5}%
  \selectfont}%
\fi\endgroup%
\begin{picture}(2724,2424)(1189,-3673)
\put(3322,-2174){\makebox(0,0)[lb]{\smash{\SetFigFont{9}{10.8}{\rmdefault}{\mddefault}{\updefault}$\mu^* = \frac{y}{C}$}}}
\end{picture}

\caption{The $\beta$--function and RG flows}
\label{fig:beta}
}}


The technical restriction on the ope (\ref{eq:ope}) means that for
unitary minimal models one can only consider perturbations
by fields of type $(1,3)$.
For a long time, the obstacle to using (\ref{eq:dg}) to analyse the
space of RG flows was the absence of any formula for the structure
constant $C$ appearing in the ope (\ref{eq:ope}).
The equations that the structure constants must satisfy were known
since 1991 from the work of Cardy and Lewellen \cite{CLew1}, but it
was not until 1999 that this obstacle was removed when Runkel
found the boundary structure constants for `A' and `D' type models in
\cite{Runkel}. 
In retrospect, the solution for the A-models is quite simple to
understand, but the general situation is still rather intriguing, as
is described in the talk by J.-B. Zuber \cite{zuber-talk}.

Runkel's results for the A-series were applied to Affleck and Ludwig's
calculation by  Recknagel et al in \cite{RRSch1}.
The calculation of the ratio (\ref{eq:dg}) is itself not hard, but the
analysis of the possible boundary conditions consistent with this
ratio is rather involved, and has a surprising result.
They found the result that the value of ${\gir}$ at the perturbative
IR fixed point of the RG flow generated by the field $\phi_{(1,3)}$
away from the $(r,s)$ boundary is given by the superposition of
$\min(r,s)$ elementary boundary conditions
\be
  B_{(r,s)}\;+\;\phi_{(1,3)}
\to
 \bigoplus_{t=1}^{\min(r,s)}
 B_{(r+s+1-2t,1)}
\;.
\label{eq:rrs}
\ee
Note that each of the end-point boundary conditions is 
{\em by itself\ }stable, but that the superposition is not itself
stable against perturbations by scalar operators, as described by
Affleck \cite{Affl1}. 

Given the surprising nature of their results, it is desirable to have
an independent check, and one way to check these results is by using
TCSA, which we describe next. 

\section{The Truncated Conformal Space Approach}

This method was introduced by Yurov and Al. Zamolodchikov
in 1990 in \cite{YZam1} as a numerical method to study bulk
perturbations of conformal field theories.
It was first applied to the boundary situation by Dorey et al in
\cite{DPTWa1}.
The idea is to construct the Hamiltonian of the model on a strip of
width $R$, restrict it to a finite-dimensional subspace, and
diagonalise it numerically.

We consider a strip of width $R$ with action (\ref{eq:sbdy}) with the
boundary perturbation restricted to one edge of the strip.
If we map the strip to the upper-half-plane, the
Hamiltonian can be expressed in terms of the operators acting on the
upper-half-plane Hilbert space as
\begin{eqnarray}
\ds
  H 
= \frac{\pi}{R}
   \Big(
   (L_0 - \frac{c}{24})
&+&
   \lambda
   \left(\frac R\pi \right)^{\!2-x}
   \!\int\!\! \varphi(e^{i\theta},e^{-i\theta})\;\D\theta
\nonumber
\\
&+&
   \mu    
   \left(\frac R\pi \right)^{\!1-h}
   \!\!\phi(1)
   \;\;\Big)
\;,
\label{eq:htcsa}
\end{eqnarray}
where the combinations
$  \lambda \left(\frac R\pi \right)^{2-x} $
and
$ \mu      \left(\frac R\pi \right)^{1-h} $
are dimensionless.

The truncation of the Hilbert space to a finite dimensional space can
be easily achieved by discarding all states for which the eigenvalue
of $L_0$ is greater than some cut-off $N$, and the matrix elements of 
(\ref{eq:htcsa}) can (often) be calculated exactly on this space.
Given the truncated space, one can then simply find the eigenstates and
eigenvalues of the Hamiltonian restricted to this space numerically.
For moderate values of $R$ (including the vicinity of the IR fixed
point, with luck) this can give reasonably accurate numerical results
for many quantities. (Again see \cite{tateo-talk} for examples).


To check the predictions of Recknagel et al.\ in the case of the
tricritical Ising model it turns out to be sufficient to examine the
spectra of the truncated Hamiltonian.
For purely boundary perturbations one expects TCSA to give the
following results for the spectra:


\vskip\baselineskip

For small $R$, the UV behaviour
\[
     E_i 
\;\sim\;
     \frac\pi R\, 
     (h_i^{\hbox{\tiny (UV)}} - \frac c{24}) + \ldots
\;.
\]

For $R$ in the `scaling region',
the IR behaviour
\[
     E_i 
\;\sim\;
     f_B
\;+\;
     \frac\pi R\, 
     (h_i^{\hbox{\tiny (IR)}} - \frac c{24}) + \ldots
\;.
\]

For large $R$,
the truncation-dominated regime.
\[
     E_i 
\;\propto\;
     R^{1-h}
\;.
\]


It is important to note that there is no guarantee that the value of
$R$ for which the scaling region becomes apparent is less than the
value at which truncation errors start to dominate. At the moment
there is little analytic control over the errors in the TCSA and it is
a matter of luck whether the scaling region is accessible in any
particular model.

Furthermore the TCSA results are expressed in terms of the bare
coupling constant $\mu$, and certain quantities may become divergent
as the cutoff $N$ is removed. For example, 
for the models analysed by Recknagel et al, $h$ is close to 1 and so
$f_B$ is a divergent quantity. This means that the individual
eigenvalues of the TCSA Hamiltonian depend strongly on $N$ and it is
only the energy differences which are physical.

As an example we give the results for the spectra of the TCSA applied
to the tricritical Ising model with the particular case of the strip with
boundary conditions $(1,1)$ and $(2,2)$, with the $(2,2)$ boundary
perturbed by the field $\phi_{(1,3)}$ in the direction of the
perturbative fixed point. 
In figure \ref{fig:flow2} we plot the normalised
energy levels  
\[
  {\cal E}_i = 2 \frac{ E_i - E_0 }{E_2 - E_0}
\;\;\hbox{vs.}\;\;
  \log R
\;,
\]
so that the second excited state always has normalised 
$ {\cal E}_2 \,{=}\, 2$.
The numbers shown are the multiplicities of the corresponding UV or IR
levels.

\FIGURE{
\parbox{6cm}{
\input{flow2.tex}
\caption{The boundary $B_{(2,2)}$ perturbed by $\phi_{(1,3)}$}
\label{fig:flow2}
}}
Note that truncation errors are already affecting the higher
levels for large $R$; these errors decrease on
increasing the truncation level.

It is one of Cardy's results \cite{Card4} that the states in the
Hilbert space of the strip with boundary conditions $(1,1)$ and
$(r,s)$ are encoded by the character $\chi_{(r,s)}$. 
This means that we can read off the boundary content of the strip by
simply counting the degeneracies of states. 

For example, the UV fixed point is the boundary condition $B_{(2,2)}$
which has singular vectors at levels $4$ and $6$ (amongst others)
which would have enabled us to identify it through 
\begin{eqnarray}
&&
   q^{c/24 - h_{2,2}}
  \, \times
  \, \chi_{(2,2)}(q)
\nonumber \\
& = & 1 + q + 2 q^2 + 3 q^4 + 4 q^5 + 6 q^6 + 8 q^8 + \ldots
\nonumber \\
& = & (1 - q^4 - q^6 + \ldots)\; \varphi(q)
\;,
\label{eq:chi22}
\end{eqnarray}
where the partition generating function $\varphi$ is 
\be
  \varphi(q) = \prod_{n=1}^\infty (1-q^n)^{-1}
\;.
\label{eq:phi}
\ee
Counting the degeneracies of the states in the IR we see that indeed
they are grouped into the characters $\chi_{(1,1)}$ and
$\chi_{(3,1)}$, in agreement with the prediction of Recknagel et
al. that
\be
  B_{(2,2)} \;+\; \phi_{(1,3)}
\to
  B_{(1,1)} \;\&\; B_{(3,1)}
\;.
\label{eq:b22+}
\ee
As a second check, we can also read off the first gap, which in the IR
should be 
\be
  {\cal E}_1^{\hbox{\tiny (IR)}}
\;=\;
   h_{(3,1)} - h_{(1,1)} 
\;=\;
   3/2
\ee 
We have indicated this on figure \ref{fig:flow2} and the TCSA data
agrees very well with this prediction.

Using the results of Recknagel et al it is possible to fill in all the
flows in figure \ref{fig:m45} marked with an asterisk, and two others
by $Z_2$ symmetry. 

This leaves unchecked the flows away from the $(2,2)$ boundary
condition generated by $\phi_{(1,2)}$, the flow generated by
$\phi_{(1,3)}$ in the direction of the non-perturbative fixed point,
and the two-dimensional space of flows generated by linear
combinations of $\phi_{(1,2)}$ and $\phi_{(1,3)}$.
In these cases we are left with the TCSA as the only viable tool at
the moment.

To complete the picture of the $\phi_{(1,3)}$ flows, in figure
\ref{fig:flow3} the spectra ${\cal E}_i$ for the perturbation of the
boundary condition $B_{(2,2)}$ by $\phi_{(1,3)}$ in the opposite
direction to that in figure \ref{fig:flow2}.
\FIGURE{
\parbox{5cm}{
~\\
~\\
\input{flow3.tex}
\caption{The boundary $B_{(2,2)}$ perturbed by $\phi_{(1,3)}$}
\label{fig:flow3}
}}
It is easy to check that this is given by
\be
  B_{(2,2)} \;-\; \phi_{(1,3)}
\to
  B_{(2,1)}
\;,
\label{eq:b22-}
\ee
an example of one of the conjectures given in \cite{RRSch1}.


The other simple flows away from $B_{(2,2)}$ in the  tricritical
Ising model are generated by 
{$\phi_{3,3}\equiv \phi_{(1,2)}$}.
In this case, 
$h_{33} \,{=}\, h_{12} \,{=}\, 1/10 \,{<}\,1/2$ 
so that the constant $f_B$ is finite and physical (being the
expectation value of the perturbing field) and so we
can plot the whole spectrum. 
In figure (\ref{fig:Graph1b12.45}) we plot 
the eigenvalues of the dimensionless operator
$(R/\pi)H$ against the dimensionless variable
$\kappa \,{=}\, \mu R^{1-h_{33}}$.

\FIGURE{
\epsfysize 6.5truecm
\epsfbox{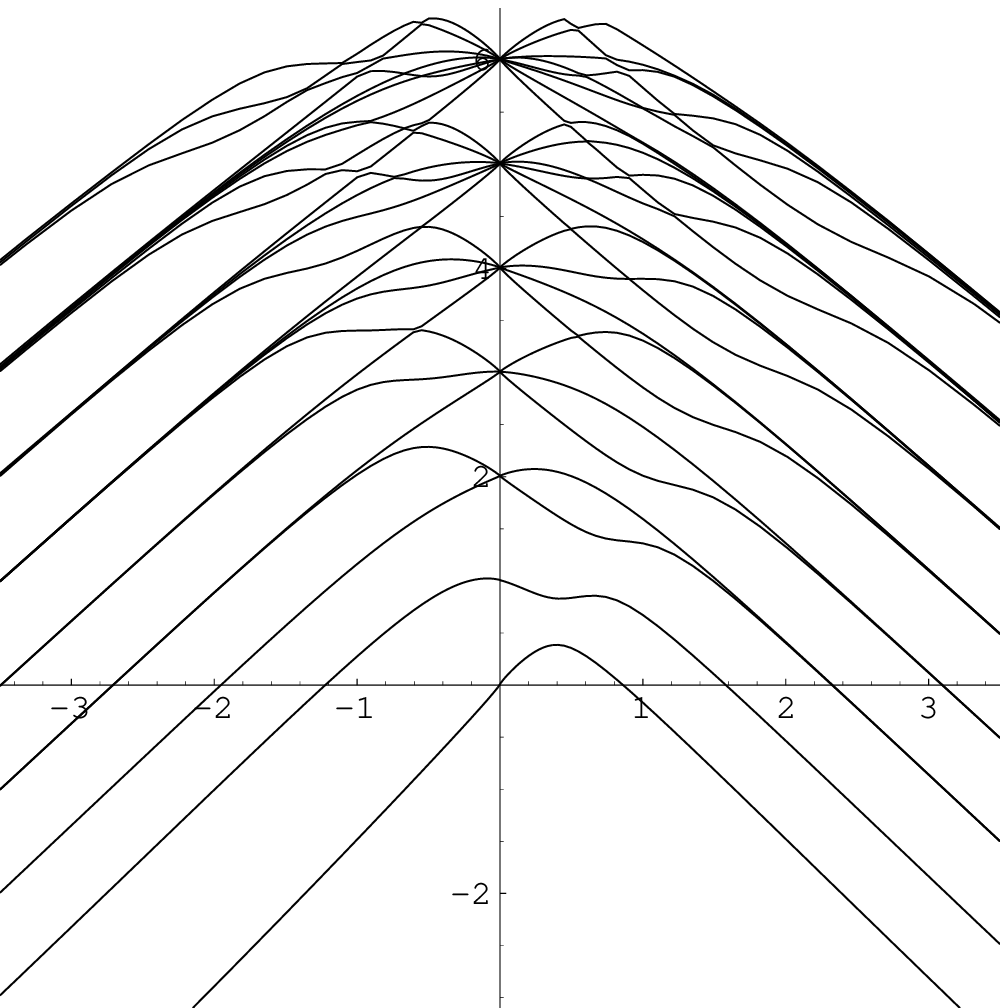}
\caption{
$M_{4,5}: B_{(22)} \pm \phi_{33}$.
\newline
The first 25
eigenvalues of $(R/\pi)H(\kappa)$ plotted vs. $\kappa$.
}
\label{fig:Graph1b12.45}
}

Here again by simply counting the states we can identify the end
points clearly as $B_{(3,1)}$ in the positive direction and
$B_{(1,1)}$ in the negative direction.
This is in agreement with the picture in figure \ref{fig:m45}, where
we have
\FIGURE{%
\noindent\parbox{4.5cm}{%
\input{m45-2.tex}
\caption{\newline{}$\phi_{(33)}$ boundary RG flows}
\label{fig:m45-2}
}}

It is also evident from the graph that there are numerous line
crossings, which are usually taken to be an indication of
integrability. 
Since the perturbing field in this case 
$\phi_{(33)}\equiv \phi_{(12)}$, this is presumably related in some
manner to the $a_2^{(2)}$ boundary affine Toda theory, in the same
manner that $\phi_{(1,3)}$ perturbations are related to the boundary
sine-Gordon theory.

Using the TCSA we can just as easily examine other models.
For unitary minimal models the relevant fields are $\phi_{(rr)}$ and
$\phi_{(r,r+2)}$. As an example, 
in figures \ref{fig:Graph1b14.67} and \ref{fig:Graph1b12.1011}
we consider for the unitary  minimal models $M_{6,7}$ and
$M_{10,11}$ the same perturbation $B_{(2,2)} \pm \phi_{(3,3)}$
as in figure \ref{fig:Graph1b12.45}.
 
\FIGURE{
\epsfysize 6.5truecm
\epsfbox{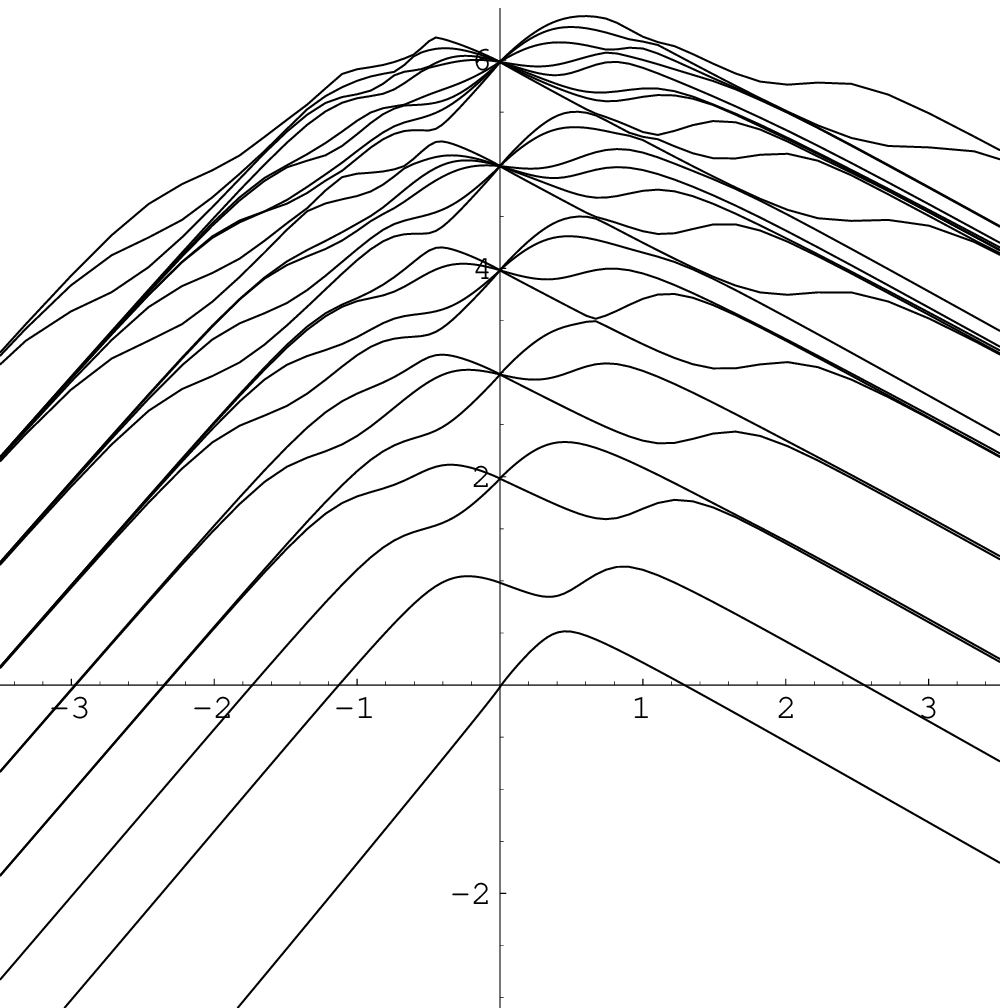}
\caption{
$M_{6,7}: B_{(22)} \pm \phi_{33}$.
\newline
The first 25
eigenvalues of $(R/\pi)H(\kappa)$ plotted vs. $\kappa$.
}
\label{fig:Graph1b14.67}
}

\FIGURE{
\epsfysize 6.5truecm
\epsfbox{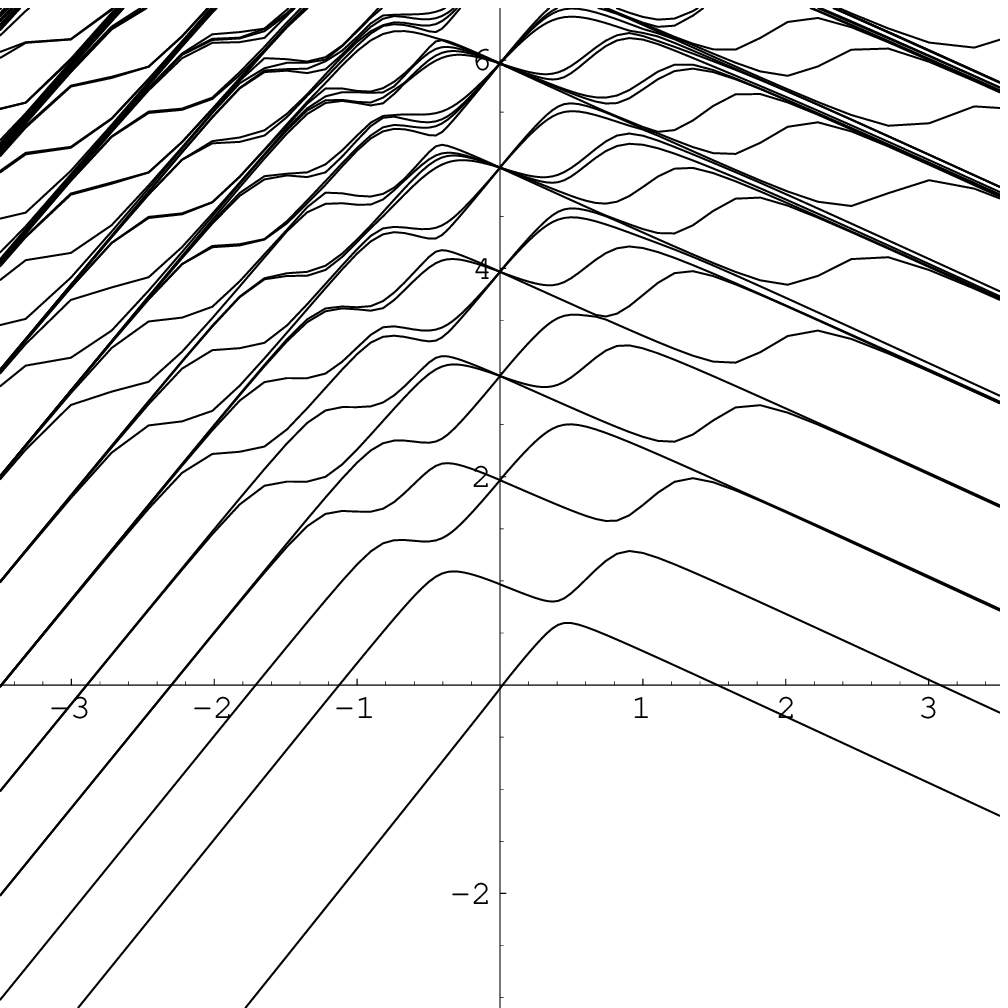}
\caption{
$M_{10,11}: B_{(22)} \pm \phi_{33}$.
\newline
The first 25
eigenvalues of $(R/\pi)H(\kappa)$ plotted vs. $\kappa$.
}
\label{fig:Graph1b12.1011}
}

It is apparent that these are deformations of figure
\ref{fig:Graph1b12.45}, and that the IR endpoints are unchanged from
$B_{(1,1)}$ and $B_{(3,1)}$, but that the line crossings so obvious
there are missing here.
However, if we continue all the way to the limiting point of the
unitary minimal models at $c\,{=}\,1$,
in figure \ref{fig:c=1} we find something quite striking.

\newpage

\FIGURE{
\epsfysize 6.5truecm
\epsfbox{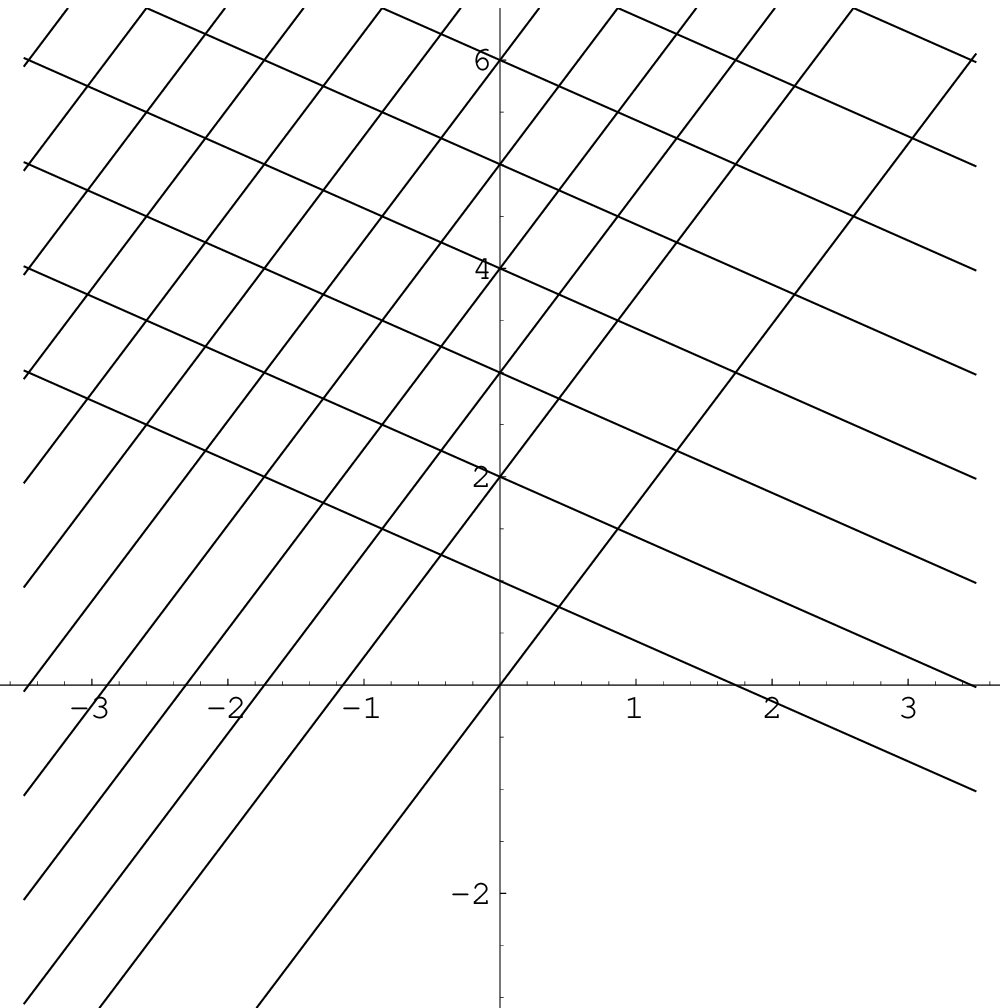}
\caption{
$M_{\infty,\infty+1}: B_{(22)} \pm \phi_{33}$.
\newline
The eigenvalues of $(R/\pi)H(\kappa)$ plotted against $\kappa$.
}
\label{fig:c=1}
}

How can we understand that the flows all become straight lines?


\noindent
The key to understanding these RG flows is the idea that

\begin{quotation}
\noindent 
\llap{``}%
perturbative flows for $c\,{<}\,1$ become identities at $c\,{=}\,1$''
\end{quotation}

\noindent
This is quite natural given the $\beta$--function for the
$\phi_{(1,3)}$ boundary perturbations of the unitary minimal model
$M_{p,p+1}$. 
In these models
\[
  c = 1 \,{-}\, \frac{6}{p(p{+}1)}
\;,\;\;\;\;
  h_{13} = 1 \,{-}\, \frac{2}{p{+}1}
\;,\;\;\;\;
  y = \frac{2}{p{+}1}
\;.
\]
The limit $c\,{=}\,1$ is given by taking $p\,{\to}\,\infty$, and the
perturbative fixed point is at $\mu^* \,{=}\, y/C \,{\to}\, 0$.
Hence as we approach $c\,{=}\,1$, the perturbative IR fixed point gets
closer and closer to the UV fixed point, until at $c\,{=}\,1$ they have
identical properties, and indeed they then represent the same boundary
condition.
However, for $c\,{<}\,1$ the UV fixed point is $B_{(2,2)}$ and the IR fixed
point the superposition $B_{(1,1)} + B_{(3,1)}$.
Hence at $c\,{=}\,1$ these must be equivalent descriptions of the same
boundary condition.

%


We need to be somewhat careful when we come to check this idea.
The ``$c\,{=}\,1$'' boundary conditions are defined as limits of the 
$c\,{<}\,1$ boundary conditions such that the space of states has a smooth
limit, i.e. given any truncation level $N$, there is some $p(N)$ such
that for $p\,{>}\,p(N)$ the space of states in $M_{p,p+1}$ up to level
$N$ has constant dimension. This requirement is already non-trivial
for $N\,{=}\,1$, as we shall see now. 

One of the fundamental results in boundary conformal field theory is
that the space of boundary fields on a boundary condition is in 1--1
correspondence with the space of states in the Hilbert space on the
upper half plane associated to that boundary condition;
furthermore, for the `A' type models considered here,
the boundary conditions $B_\alpha$ are in 1--1 correspondence with the
Virasoro representations $(\alpha)$ and the space of primary fields
interpolating two boundary conditions is given by the fusion of the
corresponding representations.
In particular, the
space of primary fields on a given boundary $B_\alpha$ is given by the
fusion product of the representation $(\alpha)$ with itself.
Hence the primary fields on the boundary condition
$B_{(2,2)}$ are  given by the fusion
\be
  (22) \times (22)
= (11) + (13) + (31) + (33)
\;.\;\;
\label{eq:fuse}
\ee
The weights (and their \clim\ limits) of these 4 primary fields are
\be
  \begin{array}{rclrl}
h_{11} &=& 0 & \to & 0 \\
h_{13} &=& 1 -  2/p + \ldots\;         & \to & 1 \\
h_{31} &=& 1 +  2/p                    & \to & 1 \\
h_{33} &=& {1}/({2 p^2})  + \ldots\;   & \to & 0 
  \end{array}
\label{eq:hs}
\ee
This immediately causes concern, as 
anyone familiar with the representation theory of the Virasoro algebra
will notice, that since $h_{33}\to 0$ as \clim, the state
\be
 L_{-1} \vec{ h_{33} }
\;,
\ee
will become a null state at $c\,{=}\,1$.
Since we require the space of states to have a smooth limit,  
we must rescale this state to obtain a {\em new} primary field of
weight 1, 
\be
   d_3
= \lim_{p\to\infty}
  \,
  \frac{1}{\sqrt{2 h_{33}}}
  \,
  L_{-1}
  \vec{h_{33}}
\;.
\label{eq:d3def}
\ee
It turns out that this field is entirely well behaved -- it might have
been the case that some of its correlation functions diverged, but
in fact they remain finite in the limit \clim.

So, in the limit \clim\ of $B_{(2,2)}$ we have (amongst other
primary fields)
\be
\begin{array}{rcl}
 2 & \hbox{ fields of weight } & 0 \\
 3 & \hbox{ fields of weight } & 1
\end{array}
\ee
These are exactly the numbers of such boundary fields that exist on a
superposition of the boundary conditions%
\footnote{
The boundary conditions $(r,1)$ and $(1,r)$ have no similar problems and
in fact have the same (smooth) limit as \clim.
}
\be
  B_{(1,1)} \;+\; B_{(3,1)}
\;.
\ee
On this superposition one expects to have (amongst other primary
fields of higher weight)
two scalar fields
\be
  \One_{(1,1)}
,\;
  \One_{(3,1)}
,
\ee
being the projectors onto the individual boundary conditions,
and three fields of weight one
\be
  \psi^{(1,1)\to(3,1)}
,\;
  \psi^{(3,1)\to(1,1)}
,\;
  \phi^{(3,1)}
,
\ee
where the fields $ \psi^{(1,1)\to(3,1)} $ and $ \psi^{(3,1)\to(1,1)} $
interpolate the two boundary conditions, while the field $
\phi^{(3,1)} $ lives on the $B_{(3,1)}$ boundary condition.

One can check \cite{GRWa1} that the combinations
\be
\begin{array}{rcl}
    \One_{(2,2)}
&=& \One_{(1,1)} \;+\; \One_{(3,1)}
\;,
\\[2mm]
    \phi_{(3,3)}
&=& \ds
    \sqrt 3\cdot \One_{(1,1)} 
    \;-\; 
    \frac{1}{\sqrt 3}\cdot \One_{(3,1)}
\;,
\end{array}
\label{eq:equivs}
\ee
of the fields on the $B_{(1,1)} + B_{(3,1)}$ boundary have exactly the
same operator product expansions as their namesakes on the $B_{(2,2)}$
boundary. Similarly one can form linear combinations of the weight one
fields on the superposition which have the same operator product
expansions with themselves and with the weight 0 fields as do those on
the single $B_{(2,2)}$ boundary.
In this way we find substantial evidence to support our conjecture
that these two boundary conditions become equal at $c\,{=}\,1$.

We can now return to our discussion of RG flows generated by
$\phi_{(3,3)}$. 
If we consider a perturbation by $\phi_{(3,3)}$, then the
corresponding Hamiltonian is
\be
  H
= H_0 \;+\;  \mu \, \phi_{(3,3)}
\;.
\ee
If we now rewrite this using (\ref{eq:equivs}), we find
\be
  H
= H_0 
  \;+\; 
   \mu 
  \Big[
  \, \sqrt 3 \cdot \One_{(1,1)}
  -
   \frac 1{\sqrt 3}\cdot \One_{(3,1)}
  \,\Big]
\;.
\ee
Hence, on states in the component boundary condition $B_{(1,1)}$
\be
  H
= H_0 \;+\; \sqrt 3 \, \mu
\;,
\ee
and on states in the component boundary condition $B_{(3,1)}$
\be
  H 
= H_0 \;-\; \frac 1{\sqrt 3} \, \mu
\;.
\ee
In other words the perturbation of $B_{(2,2)}$ by $\phi_{(3,3)}$ just
leads to the linear splitting of the energies of the states in the
$B_{(1,1)}$ and $B_{(3,1)}$ boundary conditions seen in figure
\ref{fig:c=1}. 

There is of course no such simple interpretation of the $\phi_{(3,3)}$
perturbations for $c<1$, but it appears that the identity of the
IR endpoints of the perturbation of $B_{(2,2)}$ by $\pm\phi_{(3,3)}$ 
remains unchanged all the way down to $p\,{=}\,4$ which is the smallest
value of $p$ for which this boundary perturbation exists in the
unitary minimal models.

We believe that this pattern extends to all the perturbations of `low'
weight, i.e. by fields of type $\phi_{(r,r)}$. That is these fields
can all be written at $c\,{=}\,1$ as linear combinations of projectors, and
that the IR endpoints are unchanged as one passes to the models
with $c\,{<}\,1$.

\section{Conclusions}

We have seen that the perturbations of the unitary minimal model
boundary conditions by fields of type $\phi_{(1,3)}$ can be analysed
in perturbation theory and the boundary condition of the perturbative
IR fixed point identified. In general this is a superposition of
boundary conditions.
This results can then be checked by the numerical TCSA method.

We have also seen how perturbations by fields of type $\phi_{(3,3)}$,
and in general of type $\phi_{(r,r)}$, can be understood as
deformations of especially simple RG flows at $c\,{=}\,1$.
Again these can be checked by TCSA methods.

Finally we should mention that TBA equations 
for the function $g_\alpha(L)$ have been proposed in
\cite{LSS} for the integrable perturbations
$B_{(1,r)} \pm \phi_{(1,3)}$.
So far these have not been subjected to quantitative tests, but they
agree with the IR fixed point calculated using perturbation theory
(for one direction of the flow) and with TCSA  (for the other
direction).   

This still leaves many questions open:

\vskip \baselineskip

For purely boundary flows some questions which occur are:
Can one understand any other RG flows as deformations of $c\,{=}\,1$ flows?
(Some work towards this for flows by weight one fields is contained
in \cite{GWat1}.) 
Can one find TBA systems to describe the remaining $\phi_{(1,3)}$ RG
flows with superpositions of boundary conditions at the IR endpoints?
Finally, can one find a simple picture such as the Landau-Ginzburg
picture for bulk flows which will give the qualitative pattern of
flows and fixed points?

\vskip \baselineskip

Some recent results for joint bulk and boundary perturbations will be
presented in the talk by Roberto Tateo \cite{tateo-talk}, but in general
the situation is even less clear than that for purely boundary
perturbations.

\acknowledgments

GMTW thanks Denis Bernard and Bernard Julia for the invitation to 
speak at this conference.
The work was supported in part by a TMR grant of the European
Commission, reference ERBFMRXCT960012.

%

\end{document}